\documentclass{statsoc}
\usepackage[a4paper, margin=1in]{geometry}
\usepackage{amsmath}
\usepackage{natbib}
\usepackage{amsfonts}
\usepackage{amsmath}
\usepackage{amssymb}
\usepackage{amstext}
\usepackage{amscd}
\usepackage{graphicx}%
\providecommand{\U}[1]{\protect\rule{.1in}{.1in}}

\title[Discovering Effect Modification in an Observational Study]{Discovering Effect Modification in an Observational Study of Surgical Mortality at Hospitals with Superior Nursing}

\author[K. Lee, D. S. Small, J. Y. Hsu, J. H. Silber and P. R. Rosenbaum]{Kwonsang Lee, Dylan S. Small, Jesse Y. Hsu, Jeffrey H. Silber and Paul R. Rosenbaum}\address{University of Pennsylvania, Philadelphia USA}

\begin{document}

\begin{abstract}
There is effect modification if the magnitude or stability of a treatment
effect varies systematically with the level of an observed covariate. \ A
larger or more stable treatment effect is typically less sensitive to bias
from unmeasured covariates, so it is important to recognize effect
modification when it is present. \ We illustrate a recent proposal for
conducting a sensitivity analysis that empirically discovers effect
modification by exploratory methods, but controls the family-wise error rate
in discovered groups. \ The example concerns a study of mortality and use of
the intensive care unit in 23,715 matched pairs of two Medicare patients, one
of whom underwent surgery at a hospital identified for superior nursing, the
other at a conventional hospital. \ The pairs were matched exactly for 130
four-digit ICD-9 surgical procedure codes and balanced 172 observed
covariates. \ The pairs were then split into five groups of pairs by CART in
its effort to locate effect modification. \ The evidence of a beneficial
effect of magnet hospitals on mortality is least sensitive to unmeasured
biases in a large group of patients undergoing rather serious surgical
procedures, but in the absence of other life-threatening conditions, such as a
comorbidity of congestive heart failure or an emergency admission leading to surgery.

\end{abstract}


\footnotetext{\textit{Address for correspondence}: Kwonsang Lee, Department of
Statistics, The Wharton School, University of Pennsylvania, Philadelphia, PA
19104-6340 USA, \textsf{E-mail: kwonlee@wharton.upenn.edu} 5 Nov 2016}

\section{Superior nurse staffing, surgical mortality and resource utilization
in Medicare}

\label{secIntroExample}

Hospitals vary in the extent and quality of their staffing, technical
capabilities and nursing work environments. \ Does superiority in these areas
confer benefits to patients undergoing forms of general surgery that might be
performed at most hospitals? \ To what extent and in what way do these factors
affect the cost of surgical care? \ Are they a life-saving benefit or a
pointless and unneeded expense in the case of relatively routine forms of surgery?

A recent study by Silber et al. (2016) sought to answer these questions using
Medicare data for Illinois, New York and Texas in 2004-2006. \ A useful marker
for superior staffing is superior nurse staffing, because there is a national
voluntary accreditation program to recognize excellent nursing environments,
so-called \textquotedblleft magnet hospitals\textquotedblright; see Aiken,
Havens and Sloane (2000). Additionally, it is relatively easy to use Medicare
files to determine the quantity of nurse staffing in the form of the
nurse-to-bed ratio. \ The study compared patient outcomes at 35 magnet
hospitals with nurse-to-bed ratios of 1 or more to outcomes for patients at
293 hospitals without magnet designation and with nurse-to-bed ratios less
than 1. \ For brevity, hospitals in the first group are called magnet
hospitals and those in the second group are called controls. \ The question
being asked is: How does a patient's choice of hospital, magnet or control,
affect the patient's outcomes and medical resource utilization? \ How
consequential is this choice among hospitals and what are its consequences?
\ There is no suggestion, implicit or otherwise, in this question that the
nurses are the active ingredient distinguishing magnet and control hospitals,
no suggestion that hiring nurses or changing the nurse environment would make
control hospitals perform equivalently to magnet hospitals. \ Magnet
designation marks a type of hospital, but does not identify what components
are critical in distinguishing that type of hospital. \ Indeed, the 35 magnet
hospitals had many advantages in staffing or technology: 21.5\% of magnet
hospitals were major teaching hospitals, as opposed to 5.7\% of control
hospitals; magnet hospitals had more nurses with advanced training, more
medical residents per bed, and were somewhat more likely to have a burn unit,
and to perform difficult forms of specialist surgery such as coronary bypass
surgery and organ transplantation; see Silber et al. (2016, Table 1). \ Does a
patient undergoing perhaps comparatively routine general surgery benefit from
all of these capabilities or are they wasted on such a patient?

The distinction in the previous paragraph may be restated as follows. \ The
counter-factual under study is: What would happen to a specific patient if
that patient were treated at a hospital having the superior staffing of magnet
hospitals when compared to what would happen to this same patient if treated
at a control hospital? \ The counter-factual refers to sending the patient to
one hospital or another. \ What would happen if patients were allocated to
existing hospitals in a different way? \ The counter-factual does not
contemplate changing the staffing at any hospital. \ Beds in hospitals with
superior staffing are in limited supply, and it is a matter of considerable
public importance that this limited resource be allocated to the patients most
likely to benefit from it. \

Some patients are in relatively good health and require relatively routine
care; perhaps these patients receive little added benefit from magnet
hospitals. \ Some patients are gravely ill and have poor prospects no matter
what care is provided; perhaps these patients also receive little added
benefit from magnet hospitals. \ In contrast, some patients would have poor
outcomes with inferior care and would have better outcomes with superior care;
perhaps these patients benefit most from treatment in a magnet hospital. \ Are
magnet hospitals more effective for some types of patients than for others?
\ This is the question of effect modification in our title.

Silber et al. (2016) created 25,752 matched pairs of two patients, one
undergoing general surgery at a magnet hospital, the other at a control
hospital. \ The two patients in a pair underwent the same surgical procedure
as recorded in the 4-digit ICD-9 classification of surgical procedures, a
total of 130 types of surgical procedure. \ Additionally, the matching
balanced a total of 172 pretreatment covariates describing the patient's
health prior to surgery; see Silber et al. (2016, Table 2). \ Overall, Silber
et al found significantly lower mortality at magnet hospitals than at control
hospitals (4.8\% versus 5.8\%, McNemar $P$-value $<$ 0.001), substantially
lower use of the intensive care unit or ICU (32.9\% versus 42.9\%) and
slightly shorter length of stay; see Silber et al. (2016, Table 3) where costs
and Medicare payments are also evaluated. \ Magnet hospitals had lower
mortality rates while making less use of an expensive resource, the ICU.

In one analysis, Silber et al. (2016) grouped matched pairs based on an
estimated probability of death that was controlled by the matching algorithm.
\ The lowest risk patients appeared to benefit least from magnet hospitals.
\ In contrast, the fourth quintile of risk --- a high, but not the highest,
quintile of risk --- had both lower mortality and lower cost in magnet
hospitals, whereas the highest risk quintile had lower mortality but higher
cost at magnet hospitals. \ In brief, Silber et al. (2016) found evidence of
effect modification.

Patients with very different medical problems may have similar probabilities
of death. \ It is interesting that the effect of magnet hospitals appears to
vary with patient risk, but it would be more interesting still to unpack
patient risk into its clinical constituents, and to understand how the effect
varies with these constituents. \ Clinicians do not think of patients in terms
of their probability of death, but rather in terms of their specific health
problems that are aggregated by the probability of death. \ In that sense, the
examination of effect modification in Silber et al. (2016) is too limited to
guide practice.

The current paper uses a recently proposed exploratory technique to unpack
effect modification, combined with a confirmatory technique that examines the
sensitivity of these conclusions to unmeasured biases. \ Is the ostensible
effect larger, more stable or more insensitive to unmeasured bias for certain
surgical procedure clusters or certain categories of patients defined by other
health problems?

\section{Review of effect modification in observational studies}

\label{secReview}

\subsection{Notation for causal effects, nonrandom treatment assignment,
sensitivity analysis}

In observational studies, it is known that certain patterns of treatment
effects are more resistant than others to being explained away as the
consequence of unmeasured biases in treatment assignment; see, for instance,
Rosenbaum (2004), Zubizarreta et al. (2013), Stuart and Hanna (2013).

Effect modification occurs when the size of a treatment effect or its
stability varies with the level of a pretreatment covariate, the effect
modifier. \ Effect modification affects the sensitivity of ostensible
treatment effects to unmeasured biases. \ Other things being equal, larger or
more stable treatment effects are insensitive to larger unmeasured biases; see
Rosenbaum (2004, 2005). \ As a consequence, discovering effect modification
when it is present is an important aspect of appraising the evidence that
distinguishes treatment effects from potential unmeasured biases, a concern in
every observational study. In particular, Hsu et al. (2013, 2015) discuss
sensitivity analysis in observational studies with potential effect
modification, and \S \ref{secReview} is a concise summary. \ Chesher (1984),
Crump et al. (2008), Lehrer, Pohl and Song (2015), Lu and White (2015), Wager
and Athey (2015), Athey and Imbens (2016), and Ding et al. (2016), discuss
effect modification from a different perspective, placing less emphasis on its
role in confirmatory analyses that distinguish treatment effects from
unmeasured biases in observational studies. \

There are $I$ matched pairs, $i=1,\ldots,I$, of two subjects, $j=1,2$, one
treated with $Z_{ij}=1$, the other control with $Z_{ij}=0$, so $Z_{i1}%
+Z_{i2}=1$ for each $i$. \ Subjects are matched for an observed covariate
$\mathbf{x}_{ij}$, so $\mathbf{x}_{i1}=\mathbf{x}_{i2}=\mathbf{x}_{i}$, say,
for each $i$, but may differ in terms of a covariate $u_{ij}$ that was not
measured. \ Each subject has two potential responses, $r_{Tij}$ if treated,
$r_{Cij}$ if control, exhibiting response $R_{ij}=Z_{ij}\,r_{Tij}+\left(
1-Z_{ij}\right)  \,r_{Cij}$, so the effect caused by the treatment,
$r_{Tij}-r_{Cij}$ is not seen from any subject; see Neyman (1923) and Rubin
(1974). \ Fisher's (1935) null hypothesis $H_{0}$ of no treatment effect
asserts $r_{Tij}=r_{Cij}$ for all $i$, $j$. \ Simple algebra shows that the
treated-minus-control pair difference in observed responses is $Y_{i}=\left(
Z_{i1}-Z_{i2}\right)  \left(  R_{i1}-R_{i2}\right)  $ which equals $\left(
Z_{i1}-Z_{i2}\right)  \left(  r_{Ci1}-r_{Ci2}\right)  =\pm\left(
r_{Ci1}-r_{Ci2}\right)  $ if Fisher's hypothesis $H_{0}$ is true. \ Write
$\mathcal{F}=\left\{  \left(  r_{Tij},\,r_{Cij},\,\mathbf{x}_{ij},\,u_{ij}
\right)  ,\,i=1,\ldots,I,\,j=1,2\right\}  $ for the potential responses and
covariates, and write $\mathcal{Z}$ for the event that $Z_{i1}+Z_{i2}=1$ for
each $i$.

In a randomized experiment, $Z_{i1}=1-Z_{i2}$ is determined by $I$ independent
flips of a fair coin, so $\pi_{i}=\Pr\left(  \left.  Z_{i1}=1\,\right\vert
\,\mathcal{F},\mathcal{Z}\right)  =\frac{1}{2}$ for each $i$, and this becomes
the basis for randomization inferences, for instance for tests of Fisher's
null hypothesis or for confidence intervals or point estimates formed by
inverting hypothesis tests. \ A randomization inference derives the null
distribution given $\left(  \mathcal{F},\mathcal{Z} \right)  $ of a test
statistic as its permutation distribution using the fact that the $2^{I}$
possible values of $\mathbf{Z=}\left(  Z_{i1},Z_{i2},\ldots,Z_{I2}\right)  $
each have probability $2^{-I}$ in a randomized paired experiment; see Fisher
(1935), Lehmann and Romano (2005, \S 5), or Rosenbaum (2002a, \S 2). \ A
simple model for sensitivity analysis in observational studies says that
treatment assignments in distinct pairs are independent but bias due to
nonrandom treatment assignment may result in $\pi_{i}$ that deviate from
$\frac{1}{2}$ to the extent that $1/\left(  1+\Gamma\right)  \leq\pi_{i}%
\leq\Gamma/\left(  1+\Gamma\right)  $ for $\Gamma\geq1$, and the range of
possible inferences is reported for various values of $\Gamma$ to display the
magnitude of bias that would need to be present to materially alter the
study's conclusion; see, for instance, Rosenbaum (2002a, \S 4.3.2; 2002b) for
the case of matched binary responses, as in the current paper. \ For instance,
a sensitivity analysis may report the range of possible $P$-values or point
estimates that are consistent with the data and a bias of at most $\Gamma$ for
several values of $\Gamma$.

For various approaches to sensitivity analysis in observational studies, see
Cornfield et al. (1959), Gastwirth (1992), Gilbert et al. (2003), Egleston et
al. (2009), Hosman et al. (2010) and Liu et al. (2013).

\subsection{Three strategies examining effect modification}

\label{ss3strategies}

There is effect modification if the magnitude of the effect, $r_{Tij}-r_{Cij}%
$, varies systematically with $\mathbf{x}_{i}$. \ Let $\mathcal{G}$ be a
subset of the values of $\mathbf{x}$, and define the null hypothesis
$H_{\mathcal{G}}$ to be Fisher's null hypothesis for individual $j$ in set $i$
with $\mathbf{x}_{ij}\in\mathcal{G}$, so $H_{\mathcal{G}}$ asserts that
$r_{Tij}=r_{Cij}$ for all $ij$ with $\mathbf{x}_{ij}=\mathbf{x}_{i}%
\in\mathcal{G}$. Let $\mathfrak{g}=\left(  \mathcal{G}_{1},\ldots
\mathcal{G}_{G}\right)  $ be a mutually exclusive and exhaustive partition of values of $\mathbf{x}_{ij}=\mathbf{x}_{i}$, so each pair $i$
has an $\mathbf{x}_{i}$ contained in exactly one $\mathcal{G}_{g}$. \ A simple
form of effect modification occurs if $H_{\mathcal{G}_{g}}$ is true for some
$g$ but not for other $g$. \ Write $I_{g}$ for the number pairs with
$\mathbf{x}_{i}\in\mathcal{G}_{g}$, so $I=\sum_{g=1}^{G}I_{g}$.

There are three strategies for defining the groups, $\mathfrak{g}=\left(
\mathcal{G}_{1},\ldots,\mathcal{G}_{G}\right)  $, two of which are practically
useful but technically straightforward, the third having interesting technical
aspects that we illustrate using the Medicare example.\ One useful strategy
defines the groups, $\mathfrak{g}=\left(  \mathcal{G}_{1},\ldots
,\mathcal{G}_{G}\right)  $, a priori, without reference to data.\ For example,
on the basis of clinical judgement, one might believe certain surgical
procedures are more challenging or hazardous than others, and therefore divide
the exactly matched procedures into a few groups based on clinical judgement
alone. \ Alternatively, clinical judgement might separate patients with severe
chronic conditions unrelated to the current surgery, such as congestive heart failure.

A second strategy uses an external source of data to define the groups. \ In
particular, Silber et al. (2016) fit a logit model to an external data source,
predicting mortality from covariates, $\mathbf{x}_{ij}$, then formed five
groups $\mathfrak{g}=\left(  \mathcal{G}_{1},\ldots,\mathcal{G}_{5}\right)  $
based on this predicted risk for a given $\mathbf{x}$. \ This approach made no
use of the mortality experience of the patients in the current study in
defining the groups. \ A variant of the second strategy is to split one data
set at random into two parts, create the groups using the first part, then
analyze only the second part with these, again, externally determined groups.

In both of the first two strategies, the groups, $\mathfrak{g}=\left(
\mathcal{G}_{1},\ldots,\mathcal{G}_{G}\right)  $, were determined by events
external to the outcomes reported study. \ The second strategy makes explicit
use of an external source of data, while the first strategy uses judgement
that is presumably informed historically by various external sources of data.
\ The key element in both strategies is that the groups were fixed before
examining outcomes in the current study, and in that sense are unremarkable as
groups, requiring no special handling because of their origin. \ With a priori
groups, we could use any of a variety of methods to test the $G$ hypotheses
$H_{\mathcal{G}_{g}}$ in such a way as to strongly control the family-wise
error rate at $\alpha$, meaning that the chance of falsely rejecting at least
one true $H_{\mathcal{G}_{g}}$ is at most $\alpha$ no matter which hypotheses
are true and which are false.

The third strategy that we illustrate here creates the groups, $\mathfrak{g}%
=\left(  \mathcal{G}_{1},\ldots,\mathcal{G}_{G}\right)  $, by exploratory
techniques using all of the current data, and then goes on to perform an
analysis of the same data as if the groups had been determined a priori. The
third strategy is designed so that it controls the family-wise error rate in a
sensitivity analysis despite the data-dependent generation of $G$ particular
groups from among the infinitely many ways of splitting the space of values of
the observed covariates $\mathbf{x}$. \ This strategy is discussed in detail
in Hsu et al. (2015) and it entails certain restrictions on the way the groups
are constructed.

A simple version of the strategy regresses $\left\vert Y_{i}\right\vert = \left\vert \left(
Z_{i1}-Z_{i2}\right)  \left(  R_{i1}-R_{i2}\right) \right\vert$ on
$\mathbf{x}_{i}$ using a form of regression that yields groups, such as CART.
\ Note that the unsigned $\left\vert Y_{i}\right\vert $ not the signed $Y_{i}$
are used; that is, the regression does not know who is treated and who is
control. The leaves of a CART tree become the groups, $\mathfrak{g}=\left(
\mathcal{G}_{1},\ldots,\mathcal{G}_{G}\right)  $. \ The signs of the $Y_{i}$
are then \textquotedblleft remembered,\textquotedblright\ in an analysis that
views the groups, $\mathfrak{g}=\left(  \mathcal{G}_{1},\ldots,\mathcal{G}%
_{G}\right)  $, as fixed, so it resembles analyses that would have been
appropriate with an a priori grouping of the type created by the first two strategies.

It is important to understand what is at issue in the third strategy; see Hsu
et al. (2015) for a precise and general technical discussion. \ Briefly if
obscurely, the groups, $\mathfrak{g}=\left(  \mathcal{G}_{1},\ldots
,\mathcal{G}_{G}\right)  $, and hence the hypotheses, $H_{\mathcal{G}_{g}}$,
are not stable. \ If the observed data had been slightly different, the CART
tree would have been different, and we would be testing different hypotheses.
\ What does it mean to speak about the probability of falsely rejecting
$H_{\mathcal{G}_{g}}$ if most data sets would not lead us to test
$H_{\mathcal{G}_{g}}$?

Consider the simplest case, a paired randomized experiment. \ If Fisher's null
hypothesis of no effect of any kind were true, then $Y_{i}=\left(
Z_{i1}-Z_{i2}\right)  \left(  r_{Ci1}-r_{Ci2}\right)  =\pm\left(
r_{Ci1}-r_{Ci2}\right)  $ and, given $\left(  \mathcal{F},\mathcal{Z}\right)
$, different random assignments $Z_{ij}$ always yield $\left\vert
Y_{i}\right\vert =\left\vert r_{Ci1}-r_{Ci2}\right\vert $, so all $2^{I}$
random assignments produce the same CART tree and the same $\mathfrak{g}%
=\left(  \mathcal{G}_{1},\ldots,\mathcal{G}_{G}\right)  $. \ In other words,
under $H_{0}$, the CART tree and hence $\mathfrak{g}=\left(  \mathcal{G}%
_{1},\ldots,\mathcal{G}_{G}\right)  $ is a function of $\left(  \mathcal{F}%
,\mathcal{Z}\right)  $ and not of $\mathbf{Z}$. \ Therefore, under $H_{0}$,
the $2^{I_{g}}$ possible treatment assignments for the $I_{g}$ pairs with
$\mathbf{x}_{i}\in\mathcal{G}_{g}$ each have probability $2^{-I_{g}}$,
resulting in conventional permutation tests within each of the $G$ groups,
tests that are conditionally independent given $\left(  \mathcal{F}%
,\mathcal{Z}\right)  $\ under $H_{0}$. \ The problem occurs because we are
interested in testing not just $H_{0}$, but also individual $H_{\mathcal{G}%
_{g}}$ when $H_{0}$ is false because some individuals are affected by the
treatment. \ If $H_{0}$ is false, different random assignments $\mathbf{Z}$
yield different $\left\vert Y_{i}\right\vert $, hence different CART trees and
different hypotheses, $\mathfrak{g}=\left(  \mathcal{G}_{1},\ldots
,\mathcal{G}_{G}\right)  $. \ With a bit of care, it is possible to
demonstrate two useful facts. \ First, if $r_{Tij}-r_{Cij}=0$ for all $ij$
with $\mathbf{x}_{i}\in\mathcal{G}_{g}$, then the conditional distribution
given $\mathfrak{g}=\left(  \mathcal{G}_{1},\ldots,\mathcal{G}_{G}\right)  $
and $\left(  \mathcal{F},\mathcal{Z}\right)  $ of the corresponding $Z_{ij}$
with $\mathbf{x}_{i}\in\mathcal{G}_{g}$ is its usual randomization
distribution. \ In that sense, the instability of the tree over repeated
randomizations has not distorted this conditional distribution of treatment
assignments in groups with no treatment effect. \ Second, if a method is
applied to test the $H_{\mathcal{G}_{g}}$ that would strongly control the
family-wise error rate at $\alpha$ with a priori fixed groups, then
conditionally given $\mathfrak{g}=\left(  \mathcal{G}_{1},\ldots
,\mathcal{G}_{G}\right)  $ and $\left(  \mathcal{F},\mathcal{Z}\right)  $, the
method will reject at least one null group with probability at most $\alpha$.
\ These two facts are extended to include sensitivity analyses in
observational studies and are proved as Propositions 1 and 2 of Hsu et al.
(2015). \ That paper also presents some reasons to hope that subsets of
$\mathbf{x}_{i}$ that systematically predict $\left\vert Y_{i}\right\vert $
may identify groups in which either the magnitude of $r_{Tij}-r_{Cij}$ or its
stability varies with $\mathbf{x}_{i}$.

In the current paper, we present a practical application of this third strategy.

\section{Discovering and using effect modification in the Magnet hospital
study}

\subsection{Forming groups of pairs for consideration as possible effect
modifiers}

The analyses here first broke and then re-paired the pairs in Silber et al.
(2016) so that: (i) as in Silber et al., every pair was exactly matched for
the 130 four-digit ICD-9 surgical procedure codes, (ii) the maximum number of
pairs were exactly matched for an indicator of age greater than 75, congestive
heart failure (CHF), emergency admission or not, and chronic obstructive
pulmonary disease (COPD). \ Because identically the same people were paired
differently, the balancing properties of the new pairs are exactly the same as
reported by Silber et al. (2016, Table 2), because balancing properties refer
to marginal distributions of covariates and do not depend upon who is paired
with whom.

Using \texttt{rpart} in \texttt{R}, the CART tree was built using the 22,622
pairs that were exactly matched in the sense described in the previous
paragraph, regressing $\left\vert Y_{i}\right\vert $ on $\mathbf{x}_{i}$,
where $Y_{i}$ records the difference in binary indicators of mortality. \ So,
the tree is essentially trying to locate pairs discordant for mortality,
$\left\vert Y_{i}\right\vert =1$, on the basis of exactly matched covariates.
Here, a pair is discordant if exactly one patient in the pair died within
30-days. \ CART was not offered all 130 exactly matched surgical procedure
codes, but rather 26 mutually exclusive clusters of the 130 surgical
procedures, as listed in Table 1, plus the binary covariates age$>$75, CHF,
emergency admission, and COPD. \ The resulting tree is depicted in Figure 1.
\ A few procedure clusters --- e.g., liver procedures --- are diverse, perhaps
meriting further subdivision that we do not consider here. \

The tree in Figure 1 did not use two variables that were considered by it,
namely `age$>$75' and COPD. \ In going to 22,622 pairs, we had omitted some
pairs because they were not exactly matched for `age$>$75' and COPD, but that
omission is not needed to use the tree in Figure 1.  Can we recover some of
these omitted pairs? \ To recover omitted pairs, we followed the tactic in Hsu
et al. (2015).  Specifically, we re-paired as many of the pairs that were not
used to build the tree to be exact for the 130 procedures plus CHF and
emergency admission, adding these additional 1,093 pairs to the groups in
Figure 1, making 23,715 pairs in total, or 95\% of the original study. \ All
analyses that follow refer to these 23,715 pairs.

Consider the tree in Figure 1, starting from its root at the top of the
figure. \ The tree split the population into two groups, patients without
congestive heart failure (CHF) and patients with CHF, a serious comorbid
condition. \ It then split this divided population by grouping the 26 surgical
procedure clusters. \ There are, of course, many way to group 26 procedure
clusters; for instance, there are $2^{26}-1=67,108,863$ ways to split them
into two groups. \ The procedure groups, proc1 and proc2, for patients without
CHF are slightly different from the procedure groups, proc3 and proc4, for
patients with CHF.  Table 1 displays CART's grouping of the 26 procedure
clusters into proc1, proc2, proc3 and proc4. \ In Figure 1, CART further
divided proc2 into two subsets of patients, those admitted as emergencies and
the remaining nonemergent patients. \ In Table 1, notice that proc1 and proc3
overlap extensively, as do proc2 and proc4. To the clinical eye, with a few
raised eyebrows, the procedures in proc2 and proc4 look riskier or more
complex than those in proc1 and proc3. \ Some procedure clusters slipped.
\ Appendectomy is grouped with the less risky procedures if the patient does
not have CHF, but it grouped with the more risky procedures for a patient with
CHF; however, it is unclear whether that slip is a profound insight or a hiccup.

Up to this point in the analysis, the signs of the $Y_{i}$'s for discordant
pairs have not been used; the tree knew nothing about who lived and who died
in pairs discordant for mortality.

\subsection{Informal examination of outcomes}

In \S \ref{ssPlannedAnalysis}, an analysis of mortality is carried out as
proposed in Hsu et al. (2015). \ This analysis is easier to understand if we
take a quick look first. \ The upper part of Table~2 describes mortality
informally. \ The first three numeric rows of Table~2 describe information
that CART could use in building the tree, namely the number of pairs, the
number of discordant pairs, and the proportion of discordant pairs. \ In
Table~2, 43\% = 10127/23715 of pairs are in the group 1, that is, patients
without CHF undergoing less risky procedures. \ Expressed differently, group 1
has the most pairs and the fewest discordant pairs of the five groups. \ As
one might expect given the information that CART was permitted to use, the
proportion of discordant pairs varies markedly among the groups CART built.

The next three numeric rows of Table~2 display outcomes by treatment group,
making use of $Y_{i}$ and not just $\left\vert Y_{i}\right\vert $. \ The
mortality rates for magnet and control groups are given, as is the odds ratio
computed from discordant pairs; see Cox (1970). \ All of the odds ratios are
greater than or equal to 1, suggesting higher mortality at control hospitals.
\ The largest odds ratio is in group 2, 1.53, while the largest difference in
mortality rates is in group 5, 18.6\%-16.5\% = 2.1\%. \ The odds ratio closest
to 1 is in group 3, the group most similar to group 2 except for admission
through the emergency room.

\subsection{Structured analysis of outcomes in discovered groups}

\label{ssPlannedAnalysis}

The structured analysis in Hsu et al. (2015) starts by computing randomization
tests and upper sensitivity bounds on $P$-values for each of the five groups
separately. \ In Table~2, these are based on a test of the McNemar type,
essentially binomial calculations using discordant pairs; see Cox (1970) for
discussion of paired binary data, and see Rosenbaum (2002, \S 4.3.2) for the
sensitivity analysis. \ In the bottom part of Table~2 are upper bounds on
one-sided $P$-values testing no treatment effect in a group in the presence of
a bias in treatment assignment of at most $\Gamma$.  Also given in Table~2 
are the odds ratios from discordant pairs associated with McNemar's test.

The final column in the bottom of Table ~2 gives the $P$-value for the
truncated product of $P$-values as proposed by Zaykin et al. (2002). \ The
truncated product generalizes Fisher's method for combining independent
$P$-values: the test statistic is the product of those $P$-values that are
smaller than a threshold, $\tau$, where $\tau=0.1$ in Table~2. \ Zaykin et al.
(2002) determined the null distribution of the truncated product statistic.
\ Hsu et al (2013) show that the same null distribution may be used to combine
upper bounds on $P$-values in a sensitivity analysis for a tree like Figure 1,
and that it often has superior power in this context compared to Fisher's
product of all $P$-values, essentially because sensitivity analyses promise
$P$-values that are stochastically larger than uniform for a given $\Gamma$.
\ Truncation eliminates some very large upper bounds on $P$-values.

Hsu et al. (2015) combine the truncated product statistic with the closed
testing procedure of Marcus et al. (1976) to strongly control the family-wise
error rate at $\alpha$ in a sensitivity analysis with a bias of at most
$\Gamma$. Given $G$ hypotheses, $H_{\mathcal{G}_{g}}$, $g=1,\ldots,G $,
asserting no effect in each of $G$ groups, closed testing begins by defining
$2^{G}-1$ intersection hypotheses, $H_{\mathcal{L}}$, where $\mathcal{L}%
\subseteq\left\{  1,\ldots,G\right\}  $ is a nonempty set, and $H_{\mathcal{L}%
}$ asserts that $H_{\mathcal{G}_{\ell}}$ is true for every $\ell\in
\mathcal{L}$. \ Closed testing rejects $H_{\mathcal{L}}$ if and only if the
$P$-value testing $H_{\mathcal{K}}$ is $\leq\alpha$ for every $\mathcal{K}%
\supseteq\mathcal{L}$. \ The $P$-value testing $H_{\mathcal{K}}$ is based on
the truncated product of $P$-values for $H_{\mathcal{G}_{k}}$ for
$k\in\mathcal{K}$.

The $P$-value in the final column of Table~2 tests Fisher's hypothesis $H_{0}%
$, or $H_{\mathcal{L}}$ with $\mathcal{L}=\left\{  1,2,3,4,5\right\}  $. \ For
$\Gamma=1$, this test combines five McNemar tests using the truncated product,
and in the absence of bias, the hypothesis $H_{0}$ is rejected with a
one-sided $P$-value of $2.7\times10^{-6}$. \ To complete closed testing of
subhypotheses, one performs $2^{5}-1=31$ tests of intersection hypotheses.
\ Hypothesis $H_{\left\{  3,4\right\}  }$ has a $P$-value using the truncated
product of 0.080, so neither $H_{\mathcal{G}_{3}}$ nor $H_{\mathcal{G}_{4}}$
is rejected at the 0.05 level by closed testing, but $H_{\mathcal{G}_{1}}$,
$H_{\mathcal{G}_{2}}$ and $H_{\mathcal{G}_{5}}$ are rejected. \ In short, in
the absence of bias, $\Gamma=1$, the hypothesis of no effect is rejected in
groups 1, 2, and 5.

At $\Gamma=1.05$, Fisher's hypothesis of no effect at all is rejected at the
$9.0\times10^{-5}$ level, and closed testing rejects both $H_{\mathcal{G}_{1}%
}$ and $H_{\mathcal{G}_{2}}$ at the 0.05 level. \ At $\Gamma=1.10$, Fisher's
hypothesis $H_{0}$ of no effect is rejected at the 0.012 level, but only
$H_{\mathcal{G}_{2}}$ is rejected at the 0.05 level. \ At $\Gamma=1.17$,
Fisher's hypothesis $H_{0}$ of no effect is rejected at the 0.044 level, no
individual subgroup hypothesis is rejected at the 0.05 level, but $H_{\left\{
1,2\right\}  }$ is rejected at the 0.05 level. \ At $\Gamma=1.18$, no hypothesis is
rejected at the 0.05 level.

A bias of $\Gamma=1.17$ corresponds with an unobserved covariate that doubles
the odds of having surgery at a control hospital and increases the odds of
death by more than 60\%. \ That is, stated technically, $\Gamma=1.17 $
amplifies to $\left(  \Lambda,\Delta\right)  =\left(  2.0,1.61\right)  $; see
Rosenbaum and Silber (2009). \ McNemar's test applied to all 23,715 pairs
yields a $P$-value bound of 0.063 at $\Gamma=1.15$, so this overall test is
slightly more sensitive to unmeasured biases and provides no information about subgroups.

\subsection{Use of the intensive care unit (ICU)}

In Table\ 2, magnet hospitals exhibited lower mortality than control hospitals
for ostensibly similar patients undergoing the same surgical procedure, that
is, magnet hospitals exhibited better quality. \ Does better quality cost
more? \ For resources that are allocated by a market mechanism --- say,
restaurants or hotels --- we expect better quality to cost more, but market
forces play little role in Medicare payments. \ In the absence of market
forces, it is an open question whether better quality costs more. \ Silber et
al. (2016) examine this issue in several ways, but Table\ 3 restricts
attention to the consumption of a particularly expensive resource, namely use
of the intensive care unit or ICU. \ In a hospital with inadequate nursing
staff, a patient may be placed in the ICU to ensure that the patient is
monitored, while in a hospital with superior nursing this same patient might
remain in a conventional hospital room. \ This is one mechanism by which
better quality --- lower mortality rates --- might cost less, not more.

Is the lower mortality in magnet hospitals associated with greater use of the
ICU? \ Apparently not. \ Overall and in all five groups in Figure 1, the use
of the ICU in Table~3 is lower at magnet hospitals than at control hospitals.
\ The odds ratio is largest in group 2, but it is not small in any group. \ In
various other ways also, Silber et al. (2016) found that costs were lower at
hospitals with superior nursing, despite lower mortality rates.

The closed testing procedure applied to the sensitivity analysis in the bottom
part of Table~3 rejects the null hypothesis of no effect on ICU utilization in
all five groups providing the bias in treatment assignment is at most
$\Gamma=1.5$. \ Using the method in Rosenbaum and Silber (2009), a bias of
$\Gamma=1.5$ corresponds with an unobserved covariate that increases the odds
of surgery at a control hospital by a factor of 4 and increases the odds of
going to the ICU by a factor of 2. \ Closed testing rejects no effect only in
group 2 for $1.6\leq\Gamma\leq1.8$, and cannot reject even Fisher's $H_{0}$
for $\Gamma=1.9$. \ Detailed results for group 2 are given in Table ~4.

To emphasize a point emphasized in \S \ref{secIntroExample}, Tables~2, 3 and 4
concern the effect of going to a magnet hospital rather than a control
hospital for surgery, but they do not show the specific role of nurses in this
effect. \ It is entirely plausible that superior nurse staffing would permit
more patients to stay out of the ICU, but nothing in the data speaks to this
directly. \ The main difference between the ICU and the floor of the hospital
is the higher density, often higher quality, of the nurse staffing in the ICU.
\ A hospital with a higher nurse-to-bed ratio and superior nurse staffing may
be able to care for a seriously ill patient on the hospital floor, where some
other hospital would be forced to send the same patient to the ICU.

\subsection{Other analyses and options for analysis}

The tree in Figure 1 was built for mortality, but was used also for ICU
use.\ In an additional analysis, we applied CART to each leaf of Figure 1 to
predict unsigned discordance for ICU use. \ The two interesting aspects of
this analysis were: (i) subgroup 2 in Figure 1 was not further divided; (ii)
subgroup 5 in Figure 1 was further divided, with more evidence of an effect on
ICU use among patients in this subgroup who were not admitted through the
emergency room, a pattern analogous to subgroups 2 and 3. \ An interesting
feature of this type of analysis is that it makes mortality the primary
endpoint, as it would be in most surgical studies, so only mortality
determines the initial tree for the mortality analysis, but it permits the
secondary outcome of ICU use to affect a secondary tree.

We let CART build the groups. Any analysis that used only $\left\vert
Y_{i}\right\vert $ and $\mathbf{x}_{i}$ could be used to build the groups.
\ In saying this, we mean that the strong control of the family-wise error
rate in Hsu et al. (2015) would not be affected by revisions to the tree that
used only $\left\vert Y_{i}\right\vert $ and $\mathbf{x}_{i}$. \ Indeed, a
surgeon who did not look at $Y_{i}$ could look at Figure 1 and Table~1 and
decide to regroup some of the procedure groups. \ Perhaps the surgeon would
view some of CART's decisions as clinically unwise and would change them, or
perhaps the surgeon would prefer that proc1 and proc3 be identical, and that
proc2 and proc4 be identical. \ Indeed, the surgeon might suggest fitting the
tree again, using only $\left\vert Y_{i}\right\vert $ and $\mathbf{x}_{i}$,
but subdividing some procedure clusters, say liver procedures, that seem too
broad to be clinically meaningful. \ What is critical is that the groups are
formed using $\left\vert Y_{i}\right\vert $ and $\mathbf{x}_{i}$ without using
the sign of $Y_{i}$.

\section{Summary and discussion: Confirmatory analyses that discover larger
effects by exploratory methods}

\subsection{Summary: It is important to notice subgroups with larger treatment
effects in observational studies}

\label{ssSummary}

In an observational study of treatment effects, there is invariably concern
that an ostensible treatment effect is not actually an effect caused by the
treatment, but rather some unmeasured bias distinguishing treated and control
groups. \ Larger or more stable treatment effects are more insensitive to such
concerns than smaller or more erratic effects; that is, larger biases measured
by $\Gamma$ would need to be present to explain a large and stable treatment
effect. \ These considerations motivate an interest in effect modification in
observational studies. \ Perhaps the treatment effect is larger or more stable
in certain subgroups defined by observed covariates. \ If so, the ostensible
treatment effect in such subgroups is likely to be insensitive to larger
unmeasured biases, therefore more credible, and additionally, a larger or more stable
effect is likely to be more important clinically.

The magnet hospitals had lower mortality overall, and lower or equivalent
mortality in each of the five groups. \ However, the superior staffing of
magnet hospitals was least sensitive to unmeasured bias in our group 2,
consisting of patients undergoing relatively serious forms of surgery in the
absence of other life-threatening conditions, such as congestive heart failure
or an emergency admission leading to surgery. \ Moreover, not only were
mortality rates lower in magnet hospitals for these patients (2.5\% rather
than 3.5\%), but additionally the magnet hospitals cared for these patients
with greatly reduced use of an expensive resource, namely the intensive care
unit (ICU rate of 28.9\% rather than 43.3\%). \ Determining the cost of
hospital care for Medicare patients is not straightforward, so Silber et al.
(2016) contrasted several formulas to appraise the cost of magnet hospitals.
\ In all of these formulas, use of the ICU plays a substantial part, as does
the length of stay in the hospital. \ Regardless of which formula was used,
magnet hospitals appear to produce lower mortality either at no additional
cost or with a cost savings.

A plausible interpretation of Figure 1, Table 1 and Table 2 is that: (i)
patients in groups 2, 4 and 5 should be directed to magnet hospitals, a limited
resource; (ii) the large number of comparatively healthy patients requiring
simpler surgical procedures may go to non-magnet hospitals if space in a
magnet hospital is unavailable, (iii) patients in group 3 requiring emergency
surgery should go to the nearest hospital.

\subsection{Exploration, confirmation or prediction using regression trees}

The CART method of Breiman, Friedman, Olshen and Stone (1984) immediately
attracted the attention of clinicians, in part because its relatively compact
and coarse regression trees resemble clinical thinking. Many clinical
decisions --- e.g., which patients should go to the ICU --- are discrete
choices, and distinctions that are critical in one context may be unimportant
in another context, a pattern often suggested by a CART tree. \ For a
discussion of regression trees that emphasizes its parallel with clinical
thinking, see Zhang and Singer (2010). \ Alas, CART trees can be either
unstable or too coarse or both. \ Subsequent work found that the much finer
distinctions produced by random forests, boosting or BART improved predictions
when compared with the compact trees produced by CART; see, for instance,
Schapire, Freund, Bartlett, and Lee (1998), Breiman (2001a), and Chipman,
George and McCulloch (2010). \ Although these much finer distinctions improved
prediction, and although they have a role in improving clinical assessments,
they were no longer intelligible to humans, and consequently had limited value
in scientific publications in medical journals. \ A clinician may consider an
algorithm's assessment, but the clinician cannot transfer responsibility for
the patient to the algorithm, so the clinician needs to incorporate
intelligible considerations in her decisions. Though not a recent development,
the proverb's admonition, \textquotedblleft Happy is the [\ldots\ person
\ldots\ ] that getteth understanding,\textquotedblright\ has lost none of its
force (Proverbs 3:13, King James Version). \ For all their faults, CART's
coarse trees can be understood.

The CART method, as originally proposed, did not lend itself to conventional
inference, such as hypothesis testing, much less to simultaneous inference for
the groups it produced. \ Breiman (2001b) took pride in the gap between
methods like CART and confirmatory statistical analyses of the type published
in scientific journals, but it is at least debatable whether the gap is an
asset or a liability. \

In contrast, Hsu et al. (2015) proposed a way to use CART, or similar methods,
combining exploratory construction of groups together with a confirmatory
sensitivity analysis that controls the family-wise error rate in the
constructed groups. \ All of the data are used to build the tree and all of
the data are used in confirmatory analyses. \ This is important in Table 2
because the number of pairs discordant for mortality is not large in some
groups --- happily, most people survive surgery --- so sample splitting to
build the groups would leave less data for confirmatory analyses. \ The double
use of all of the data works by having CART predict $|Y_{i}|$ from $x_{i}$
without knowing who is treated and who is control, then using the signed
$Y_{i}$ in confirmatory analyses with CART's groups.  CART trees can be
unstable, so the tree should be regarded as an interesting partition of
the data, not a search for a ``true'' partition.  The formal hypothesis tests are
conditional inferences given CART's partition: they correctly use, but do not
endorse, the partition.

Will a tree built from $|Y_{i}|$ be useful in the study of effect
modification? \ It is straightforward to construct theoretical examples in
which an analysis of $\left\vert Y_{i}\right\vert $ would miss effect
modification that an analysis of $Y_{i}$ might find. \ Obviously, a tree built
from all of the $Y_{i}$ is preferable, but this would preclude a confirmatory
analysis using the same data. \ As noted by Hsu et al. (2015), a result of
Jogdeo (1977, Theorem 2.2) provides some encouragement. \ A simple version of
this result says: if $Y_{i}=\mu_{i}+\epsilon_{i}$, $\mu_{i}\geq0$,
$i=1,\ldots,I$, where the errors $\epsilon_{i}$ are independent and
identically distributed with a unimodal distribution symmetric about zero,
then $\left\vert Y_{i}\right\vert $ is stochastically larger than $\left\vert
Y_{j}\right\vert $ whenever $\mu_{i}>\mu_{j}$. \ Under this simple model,
trees that form groups from the level of $\left\vert Y_{i}\right\vert $ have
some hope of finding groups heterogeneous in $\mu_{i}$. \ True, if the
$\epsilon_{i}$ are not identically distributed, if the dispersion of
$\epsilon_{i}$ varies with $i$, then the groups may be affected by both level
and dispersion; however, sensitivity to unmeasured bias is also affected by
both the level and dispersion of the treatment effects, so groups reflecting
unequal dispersion are interesting also. \ For additional encouragement, see
also the simulation results in Hsu et al. (2015).

\subsection{Alternative methods}

As noted in \S \ref{ss3strategies}, there are three basic approaches to
confirmatory sensitivity analyses for effect modification. \ One approach
starts with a priori groups, or, what amounts to the same thing, groups built
from one or more external data sets. \ Essentially this approach was used in
Silber et al. (2016) for these data. \ The five groups were defined by
quintiles of risk-of-death as estimated using a model fit to another set of
data. \ That analysis was enlightening, but the plausible interpretation at
the end of \S \ref{ssSummary} makes useful distinctions that risk quintiles do
not make.

Another approach is to: (i) split the data into two parts at random, (ii) form
patient groups from $Y_{i}$ rather than $\left\vert Y_{i}\right\vert $ using
the first part of the data, (iii) discard the first part, (iv) perform a
confirmatory analysis on the second part using the patient groups formed from
the first part. \ This approach is attractive when $I$ is very large. \ For
some indirectly related theory, see Heller et al.\ (2009). \ Presumably, if we
had twice as many pairs as we actually had, $I\rightarrow2I$, if we split the
data in half as just described, then the resulting analysis would be uniformly
better than the analysis we did with half as much data, because: (i) the tree
would be better having been built from $Y_{i}$ instead of $\left\vert
Y_{i}\right\vert $, but (ii) the confirmatory analysis would have the same
quantity of data as our confirmatory analysis. \ Silber et al. (2016) used
data from New York, Illinois and Texas primarily because purchasing Medicare
data is expensive. \ There are, however, 47 more states where these came from.%

\section*{References}%
%

\setlength{\hangindent}{12pt}
\noindent
Aiken, L. H., Havens, D. S. and Sloane, D. M. (2000) The magnet nursing
services recognition program. \textit{Am. J. Nursing}, \textbf{100}, 26-35.
\ \textsf{http://www.nursecredentialing.org/Magnet/ProgramOverview.aspx}%

\setlength{\hangindent}{12pt}
\noindent
Athey, S. and Imbens, G. (2016) Recursive partitioning for heterogeneous
causal effects. \textit{Proc. Nat. Acad. Sci.}, \textbf{113}, 7353-7360.%

\setlength{\hangindent}{12pt}
\noindent
Breiman, L., Friedman, J. H., Oshen, R. A., and Stone, C. J. (1984)
\textit{Classification and Regression Trees}. \ California: Wadsworth.%

\setlength{\hangindent}{12pt}
\noindent
Breiman, L., (2001a) Random forests. \textit{Mach. Learn.}, \textbf{45}, 5-32.%

\setlength{\hangindent}{12pt}
\noindent
Breiman, (2001b) Statistical modeling: The two cultures (with Discussion).
\textit{Statist. Sci.} \textbf{16}, 199-231.%

\setlength{\hangindent}{12pt}
\noindent
Chesher, A. (1984) Testing for neglected heterogeneity. \textit{Econometrica},
\textbf{52}, 865-872.%

\setlength{\hangindent}{12pt}
\noindent
Chipman, H. A., George, E. I. and McCulloch, R. E. (2010) BART: Bayesian
additive regression trees. \textit{Ann. Appl. Statist.}, \textbf{4}, 266-298.%

\setlength{\hangindent}{12pt}
\noindent
Cornfield, J., Haenszel, W., Hammond, E., Lilienfeld, A., Shimkin, M., Wynder,
E. (1959) Smoking and lung cancer. \textit{J. Nat. Cancer Inst.}, \textbf{22}, 173-203.%

\setlength{\hangindent}{12pt}
\noindent
Cox, D. R. (1970) \textit{Analysis of Binary Data}. London: Metheun.%

\setlength{\hangindent}{12pt}
\noindent
Crump, R. K., Hotz, V. J., Imbens, G. W., \& Mitnik, O. A. (2008)
Nonparametric tests for treatment effect heterogeneity. \textit{Rev. Econ.
Statist.}, \textbf{90}, 389-405.%

\setlength{\hangindent}{12pt}
\noindent
Ding, P., Feller, A. and Miratrix, L. (2016) Randomization inference for
treatment effect variation. \textit{J. Roy. Statist. Soc.} B \textbf{78}, 655-671.%

\setlength{\hangindent}{12pt}
\noindent
Egleston, B. L., Scharfstein, D. O. and MacKenzie, E. (2009) On estimation of
the survivor average causal effect in observational studies when important
confounders are missing due to death. \textit{Biometrics}, \textbf{65}, 497-504.%

\setlength{\hangindent}{12pt}
\noindent
Fisher, R. A. (1935) \textit{The Design of Experiments}. Edinburgh: Oliver \& Boyd.%

\setlength{\hangindent}{12pt}
\noindent
Gastwirth, J. L. (1992) Methods for assessing the sensitivity of statistical
comparisons used in Title VII cases to omitted variables. \textit{Jurimetrics}%
, \textbf{33}, 19-34.%

\setlength{\hangindent}{12pt}
\noindent
Gilbert, P., Bosch, R., Hudgens, M. (2003) Sensitivity analysis for the
assessment of the causal vaccine effects on viral load in HIV vaccine trials.
\textit{Biometrics}, \textbf{59}, 531-41.%

\setlength{\hangindent}{12pt}
\noindent
Heller, R., Rosenbaum, P.R. and Small, D.S. (2009) Split samples and design
sensitivity in observational studies. \textit{J. Am. Statist. Assoc.},
\textbf{104}, 1090-1101.%

\setlength{\hangindent}{12pt}
\noindent
Hosman, C. A., Hansen, B. B. and Holland, P. W. H. (2010) The sensitivity of
linear regression coefficients' confidence limits to the omission of a
confounder. \textit{Ann. Appl. Statist.}, \textbf{4}, 849-870.%

\setlength{\hangindent}{12pt}
\noindent
Hsu, J. Y., Small, D. S. and Rosenbaum, P. R.\ (2013) Effect modification and
design sensitivity in observational studies. \textit{J. Am. Statist. Assoc.},
\textbf{108}, 135-148.%

\setlength{\hangindent}{12pt}
\noindent
Hsu, J. Y., Zubizarreta, J. R., Small, D. S. and Rosenbaum, P. R.\ (2015)
Strong control of the familywise error rate in observational studies that
discover effect modification by exploratory methods. \textit{Biometrika},
\textbf{102}, 767--782.%

\setlength{\hangindent}{12pt}
\noindent
Jogdeo, K. (1977) Association and probability inequalities. \ \textit{Ann.
Statist.}, \textbf{5}, 495-504.%

\setlength{\hangindent}{12pt}
\noindent
Lehrer, S. F., Pohl, R. V., \& Song, K. (2015) Targeting Policies: Multiple
Testing and Distributional Treatment Effects.
http://post.queensu.ca/\symbol{126}lehrers/temult.pdf%

\setlength{\hangindent}{12pt}
\noindent
Liu, W., Kuramoto, J. and Stuart, E. (2013) Sensitivity analysis for
unobserved confounding in nonexperimental prevention research. \textit{Prev.
Sci.}, \textbf{14}, 570-580.%

\setlength{\hangindent}{12pt}
\noindent
Lu, X. and White, H. (2015) Testing for treatment dependence of effects of a
continuous treatment. \textit{Econometric Theory}, \textbf{31}, 1016-1053.%

\setlength{\hangindent}{12pt}
\noindent
Marcus, R., Peritz, E. and Gabriel, K. R. (1976) On closed testing procedures
with special reference to ordered analysis of variance. \textit{Biometrika},
\textbf{63}, 655--60.%

\setlength{\hangindent}{12pt}
\noindent
Neyman, J. (1923, 1990) On the application of probability theory to
agricultural experiments. \textit{Statist. Sci.}, \textbf{5}, 463-480.%

\setlength{\hangindent}{12pt}
\noindent
Rosenbaum, P. R. (2002a) \textit{Observational Studies} (2$^{nd}$ edition) New
York: Springer.%

\setlength{\hangindent}{12pt}
\noindent
Rosenbaum, P. R. (2002b) Attributing effects to treatment in matched
observational studies. \textit{J. Am. Statist. Assoc.}, \textbf{97}, 183-192.%

\setlength{\hangindent}{12pt}
\noindent
Rosenbaum, P. R. (2004) Design sensitivity in observational studies.
\textit{Biometrika}, \textbf{91}, 153-164.%

\setlength{\hangindent}{12pt}
\noindent
Rosenbaum, P. R. (2005) Heterogeneity and causality: unit heterogeneity and
design sensitivity in observational studies. \textit{Am. Statist.},
\textbf{59}, 147-152.%

\setlength{\hangindent}{12pt}
\noindent
Rosenbaum, P. R. and Silber, J. H. (2009) Amplification of sensitivity
analysis in observational studies. \textit{J. Am. Statist. Assoc.},
\textbf{104}, 1398-1405. \ (\texttt{amplify} function in the \texttt{R}
package \texttt{sensitivitymv})%

\setlength{\hangindent}{12pt}
\noindent
Rubin, D. B. (1974) Estimating causal effects of treatments in randomized and
nonrandomized studies. \textit{J. Educ. Psych.}, \textbf{66}, 688-701.%

\setlength{\hangindent}{12pt}
\noindent
Schapire, R.E., Freund, Y., Bartlett, P. and Lee, W.S. (1998) Boosting the
margin: a new explanation for the effectiveness of voting methods.
\textit{Ann. Statist.}, \textbf{26}, 1651-1686.%

\setlength{\hangindent}{12pt}
\noindent
Silber, J. H., Rosenbaum, P. R., McHugh, M. D., Ludwig, J. M., Smith, H. L.,
Niknam, B. A., Even-Shoshan, O., Fleisher, L. A., Kelz, R. R. and Aiken, L. H.
(2016) Comparison of the value of better nursing work environments across
different levels of patient risk. \textit{JAMA Surgery}, \textbf{151}, 527-536.%

\setlength{\hangindent}{12pt}
\noindent
Stuart, E. A. and Hanna, D. B. (2013) Should epidemiologists be more sensitive
to design sensitivity?\textquotedblright\ \textit{Epidemiol.}, \textbf{24}, 88-89.%

\setlength{\hangindent}{12pt}
\noindent
Wager, S. and Athey, S. (2015) Estimation and inference of heterogeneous
treatment effects using random forests. arXiv preprint arXiv:1510.04342.%

\setlength{\hangindent}{12pt}
\noindent
Zaykin, D. V., Zhivotovsky, L. A., Westfall, P. H., and Weir, B. S. \ (2002)
Trucated product method of combining $P$-values. \textit{Genet. Epidemiol.},
\textbf{22}, 170-185. (\texttt{truncatedP} function in the \texttt{R} package
\texttt{sensitivitymv})%

\setlength{\hangindent}{12pt}
\noindent
Zhang, H. and Singer, B. H. (2010) \textit{Recursive Partitioning and
Applications}. \ New York: Springer.%

\setlength{\hangindent}{12pt}
\noindent
Zubizarreta, J. R., Cerd\'{a}, M. and Rosenbaum, P. R. (2013) Effect of the
2010 Chilean earthquake on posttraumatic stress: Reducing sensitivity to
unmeasured bias through study design. \textit{Epidemiol.}, \textbf{24}, 79-87.

\begin{table}
\caption{Grouping of procedure clusters, with and without congestive heart failure (CHF).}
\label{tabProc}
\centering
\begin{tabular}{ll|cccc}
\hline
& & No CHF & CHF & No CHF & CHF \\
& Procedure Cluster & proc1 & proc3 & proc2 & proc4 \\
\hline
1 & Adrenal procedures & x & x &   &   \\
2 & Appendectomy & x &   &   & x \\
3 & Bowel anastamoses &   &   & x & x \\
4 & Bowel procedures, other &   &   & x & x \\
5 & Breast procedures & x & x &   &   \\
6 & Esophageal procedures &   & x & x &   \\
7 & Femoral hernia procedures & x & x &   &   \\
8 & Gallbladder procedures & x & x &   &   \\
9 & Incisional and abdominal hernias & x & x &   &   \\
10 & Inguinal hernia procedures & x & x &   &   \\
11 & Large bowel resection &   &   & x & x \\
12 & Liver procedures & x &   &   & x \\
13 & Lysis of adhesions &   &   & x & x \\
14 & Ostomy procedures &   &   & x & x \\
15 & Pancreatic procedures &   & x & x &   \\
16 & Parathyroidectomy & x & x &   &   \\
17 & PD access procedure &   &   & x & x \\
18 & Rectal procedures & x & x &   &   \\
19 & Repair of vaginal fistulas & x & x &   &   \\
20 & Small bowel resection &   &   & x & x \\
21 & Splenectomy &   &   & x & x \\
22 & Stomach procedures &   &   & x & x \\
23 & Thyroid procedures & x & x &   &   \\
24 & Ulcer surgery &   &   & x & x \\
25 & Umbilical hernia procedures & x &   &   & x \\
26 & Ventral hernia repair & x & x &   &   \\
\hline
\end{tabular}
\end{table}

\begin{table}
\caption{\label{tabMortality} Mortality in 23,715 matched pairs of a patient receiving surgery at a magnet hospital or a control hospital, where the pairs have been divided into five groups selected by CART. }
\centering
\begin{tabular}{c|ccccc|c}
\hline
& \multicolumn{5}{|c|}{Subgroups} & Pooled \\
& Group 1 & Group 2 & Group 3 & Group 4 & Group5  \\ \hline
CHF & no & no & no & yes & yes \\
Procedures & proc1 & proc2 & proc2 & proc3 & proc4 \\
ER admission & both & no & yes & both & both \\
\hline
Number of Pairs & 10127 & 5636 & 2943 & 2086 & 2923 &  23715 \\
Discordant Pairs & 210 & 293 & 488 & 217 & 760 &  1968 \\
Percent Discordant \% & 2.1 & 5.2 & 16.6 & 10.4 & 26.0 &  8.3 \\
Odds Ratio & 1.41 & 1.53 & 1.09 & 1.28 & 1.18 &  1.23 \\
Morality \%, Magnet & 0.9 & 2.5 & 10.1 & 4.9 & 16.5 & 4.7  \\
Morality \%, Control & 1.3 & 3.5 & 10.8 & 6.2 & 18.6 &  5.6 \\ \hline \hline
\multicolumn{7}{c}{Sensitivity analysis: Upper bounds on $P$-values for various $\Gamma$} \\ \hline
$\Gamma$ & \multicolumn{5}{|c|}{Subgroups} &  Truncated \\
& Group 1 & Group 2 & Group 3 & Group 4 & Group 5 & Product  \\  \hline
1.00 & 0.008 & 0.000 & 0.195 & 0.039 & 0.013 & 0.000 \\
1.05 & 0.019 & 0.001 & 0.374 & 0.080 & 0.062 & 0.000 \\
1.10 & 0.042 & 0.003 & 0.576 & 0.143 & 0.184 & 0.012 \\
1.15 & 0.079 & 0.010 & 0.753 & 0.230 & 0.386 & 0.032 \\
1.17 & 0.099 & 0.015 & 0.809 & 0.270 & 0.479 & 0.044 \\
1.20 & 0.135 & 0.025 & 0.875 & 0.335 & 0.616 & 0.163 \\
\hline
\end{tabular}
\end{table}

\begin{table}
\caption{\label{tabICU} Use of the intensive care unit (ICU) in 23,715 matched pairs of a patient receiving surgery at a  magnet hospital or a control hospital, where the pairs have been divided into five groups indicated in Figure 1. }
\centering
\begin{tabular}{c|ccccc|c}
\hline
& \multicolumn{5}{|c|}{Subgroups} & Pooled \\
& Group 1 & Group 2 & Group 3 & Group 4 & Group5  \\ \hline
CHF & no & no & no & yes & yes \\
Procedures & proc1 & proc2 & proc2 & proc3 & proc4 \\
ER admission & both & no & yes & both & both \\
\hline
Number of Pairs & 10127 & 5636 & 2943 & 2086 & 2923 & 23715 \\
Discordant Pairs & 2675 & 2361 & 1282 & 859 & 970 & 8147 \\
Percent Discordant \% & 26.4 & 41.9 & 43.6 & 41.2 & 33.2 & 34.4 \\
Odds ratio & 1.63 & 2.05 & 1.67 & 1.70 & 1.88 & 1.78 \\
ICU \%, Magnet & 15.3 & 28.9 & 53.8 & 41.0 & 69.8 & 32.3 \\
ICU \%, Control & 21.7 & 43.3 & 64.6 & 51.7 & 80.0 & 42.0 \\
\hline \hline
\multicolumn{7}{c}{Sensitivity analysis: Upper bounds on $P$-values for various $\Gamma$} \\ \hline
$\Gamma$ & \multicolumn{5}{|c|}{Subgroups} &  Truncated \\
& Group 1 & Group 2 & Group 3 & Group 4 & Group 5 & Product  \\  \hline
1 & 0.000 & 0.000 & 0.000 & 0.000 & 0.000 & 0.000 \\
1.5 & 0.017 & 0.000 & 0.037 & 0.040 & 0.000 & 0.000 \\
1.6 & 0.312 & 0.000 & 0.254 & 0.203 & 0.009 & 0.000 \\
1.7 & 0.849 & 0.000 & 0.651 & 0.511 & 0.074 & 0.000 \\
1.8 & 0.993 & 0.002 & 0.916 & 0.798 & 0.276 & 0.049 \\
1.9 & 1.000 & 0.047 & 0.989 & 0.945 & 0.582 & 0.235 \\
\hline
\end{tabular}
\end{table}

\begin{table}
\caption{\label{tabGroup2} Mortality and ICU use in 5,636 pairs in Group 2.  The table counts pairs of patients, not individual patients.}
\centering
\begin{tabular}{r|rrr|r}
\hline
Control Hospital & \multicolumn{3}{|c|}{Magnet Hospital} & \\ \hline
& Dead & Alive, ICU & Alive, no ICU & Total \\ \hline
Dead & 23 & 72 & 105 & 200 \\
Alive, ICU & 60 & 744 & 1493 & 2297 \\
Alive, no ICU & 56 & 726 & 2357 & 3139 \\ \hline
Total & 139 & 1542 & 3955 & 5636 \\
\hline
\end{tabular}
\end{table}

\begin{figure}
\centering
\includegraphics[width=\textwidth]{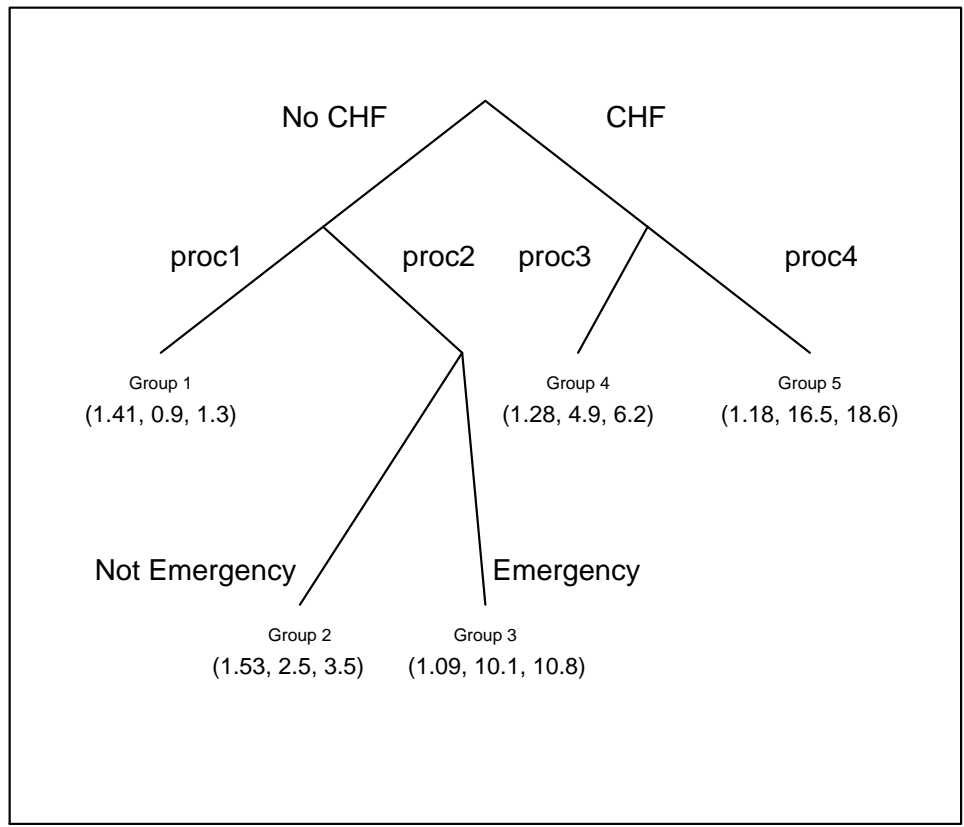}
\caption{\label{fig1}Mortality in 23,715 matched pairs of two Medicare patients, one receiving surgery at a magnet hospital identified for superior nursing, the other undergoing the same surgical procedure at a conventional control hospital. The three values (A,B,C) at the nodes of the tree are: A = McNemar odds ratio for mortality, control/magnet, B = 30-day mortality rate (\%) at the magnet hospitals, C = 30-day mortality rate (\%) at the control hospitals.}
\end{figure}

\end{document}